# Identification of excitonic phonon sideband by photoluminescence spectroscopy of single-walled carbon-13 nanotubes

Yuhei Miyauchi and Shigeo Maruyama*

Department of Mechanical Engineering, The University of Tokyo, 7-3-1 Hongo, Bunkyo-ku, Tokyo 113-8656, Japan

**Abstract**

We have studied photoluminescence (PL) and resonant Raman scatterings of single-walled carbon nanotubes (SWNTs) consisting of carbon-13 ($SW^{13}CNTs$) synthesized from a small amount of isotopically modified ethanol. There was almost no change in the Raman spectra shape for $SW^{13}CNTs$ except for a downshift of the Raman shift frequency by the square-root of the mass ratio 12/13. By comparing photoluminescence excitation (PLE) spectra of $SW^{13}CNTs$ and normal SWNTs, the excitonic phonon sideband due to strong exciton-phonon interaction was clearly identified with the expected isotope shift.

*Corresponding author.

Electronic address: maruyama@photon.t.u-tokyo.ac.jp

Recently, photoluminescence (PL) of single-walled carbon nanotubes (SWNTs) has been intensively investigated for the optical characterization of SWNTs [1-8]. By plotting PL emission intensities as a function of emission and excitation photon energy, Bachilo *et al.* [2] obtained a two-dimensional map of relative emission intensities. Hereafter, we refer to such a plot of PL spectra of SWNTs as a "PL map". A peak in a PL map corresponds to the excitation transition energy of the second subband ($E_{22}$) and the photon emission energy of the first subband ($E_{11}$) of a specific SWNT, which can be used for assignment of chiral indices (n, m) [9]. Theoretical studies and recent experiments have demonstrated that these optical transitions in SWNTs are dominated by strongly correlated electron-hole states in the form of excitons [10-13]. Since these pairs of transition energies depend on the nanotube's (n, m) structure, we can separately measure a photoluminescence excitation (PLE) spectrum from specific (n, m) SWNTs as a cross section of the PL map at an energy corresponding to the emission of the relevant SWNTs. As far as semiconducting SWNTs are concerned, such PL mapping is one of the most promising approaches for the determination of the structure distribution in a bulk SWNT sample, if combined with an appropriate theoretical estimation of relative PL intensities depending on the electronic structure specific to each (n, m) type [14, 15]. Hence, photoluminescence spectroscopy is a powerful tool not only for investigation of electronic properties of SWNTs, but in advancing toward (n, m)-controlled synthesis of SWNTs, which has never been achieved.

In a PL map, we generally find peaks other than bright PL peaks already assigned to particular nanotube species, whose origins have not been elucidated [2, 5, 6]. Recently, some PL peaks from DNA-wrapped nanotubes [16] were explained by a phonon-assisted excitonic absorption and recombination process [17]. On the other hand, theoretical prediction of the excitonic phonon sideband shape [18] was in good agreement with the PL spectrum from an individual nanotube [19]. Since these unassigned features may overlap with other PL peaks if a measured sample is a mixture of various (n, m) structures, it is very important to understand the origins of all the features in a PL map for the accurate measurement of relative PL intensities of each (n, m) nanotube. To investigate the origins of these unassigned features, we measured PL and Raman scatterings of SWNTs consisting of the carbon-13 isotope (SW$^{13}$CNTs). Due to the atomic mass difference between the $^{12}$C and $^{13}$C isotopes, phonon energies in SW$^{13}$CNTs are expected to be smaller than in normal SWNTs. Hence, PL peaks originating from exciton-phonon interactions should be clearly distinguishable from PL peaks without exciton-phonon interaction. We synthesized SW$^{13}$CNTs from 0.5 grams of isotopically modified ethanol (1,2-$^{13}$C$_2$, 99%, Cambridge Isotope Laboratories, Inc.) by the alcohol catalytic chemical vapor deposition (ACCVD) method [20, 21] optimized for the efficient production of SWNTs from a very small amount of ethanol, which is similar to the technique used for the SWNT synthesis from fullerene [22]. Further details about the synthesis of SW$^{13}$CNTs will be presented elsewhere [23].

In order to measure PL spectra from individual SWNTs in a surfactant suspension [1], the 'as-grown' material was dispersed in D$_2$O with 0.5 wt % sodium dodecylbenzene sulfonate (NaDDBS) [5] by heavy sonication with an ultrasonic processor (Hielscher GmbH, UP-400S with H3/Micro Tip 3) for 1 h at a power flux level of 460 W/cm$^2$. These suspensions were then centrifuged (Hitachi Koki himac CS120GX with S100AT6 angle rotor) for 1 h at 386 000 *g* and the supernatants, rich in isolated SWNTs, were used in the PL measurements.

Near infrared emission from the samples was recorded while the excitation wavelength was scanned from VIS to NIR range. The measured spectral data were corrected for wavelength-dependent variations in excitation intensity and detection sensitivity. The excitation and emission spectral slit widths were both 10 nm (15~30 meV for excitation and ~10 meV for emission in the measuring range), and scan steps were 5 nm on both axes. In addition to PL maps over a wide emission energy range, we obtained PLE spectra of (7, 5) nanotubes (emission at 1026.5 nm (1.208 eV)) with narrower excitation spectral slit width (5 nm: 8~15 meV in the measuring range) and smaller scan steps (2 nm) to obtain higher resolution spectra. The Raman spectra were measured using a CHROMEX 501is spectrometer, an ANDOR Technology DV401-FI CCD system, and a SEKI TECHNOTRON Corp. STR250 optical system. The photoluminescence spectra were measured with a HORIBA SPEX Fluorolog-3-11 spectrofluorometer with a liquid-nitrogen-cooled InGaAs near IR detector.

Figure 1 (a, b) compares Raman spectra for SW$^{13}$CNTs and normal SWNTs excited with a 2.54eV (488 nm) laser. The spread in the G-band peak at 1590 cm$^{-1}$ and the smaller D-band signal at 1350 cm$^{-1}$ for both SW$^{13}$CNTs and normal SWNTs suggest high quality nanotubes were synthesized. The strong radial breathing mode (RBM) peaks at around 150-300 cm$^{-1}$ were also observed for each sample. There was no intrinsic change in Raman spectra shape for SW$^{13}$CNTs, but the Raman shift frequency was $\sqrt{12/13}$ times smaller, confirming phonon energies in SW$^{13}$CNTs are $\sqrt{12/13}$ times smaller than in normal SWNTs because of the heavier carbon atoms.

Figure 1 (c, d) compares PL maps of these samples. Major peaks in PL maps are marked with (n, m) indices assigned by Bachilo *et al.* [2]. It can be confirmed that the chirality distribution of the SW$^{13}$CNT sample is almost the same as for normal SWNTs. Furthermore, detailed peak positions of major PL peaks are in good agreement between the two samples, indicating that electronic properties and the nanotube environment are virtually unchanged. Hence, phonon energies are the only principal difference between our SW$^{13}$CNTs and normal SWNTs. Since the phonon energies with the same wave vectors are different, it is expected that a PL peak with phonon excitations can be distinguished by an isotopic shift of the peak. The amount of this isotopic shift is expected to be consistent with the Raman spectroscopy results.

Fig. 2 shows high resolution PL maps for SW$^{13}$CNTs and normal SWNTs and Fig. 2c compares PLE spectra corresponding to (7, 5) nanotubes (emission at 1.208 eV) for both SW$^{13}$CNTs and normal SWNTs. As described above, some small PL peaks around the main peaks were observed as shown in Fig.2. For example, we find three small peaks above and below the main PL peak corresponding to $E_{11}$ (1.208 eV, out of measuring range) and $E_{22}$ (1.923 eV) transition energies of (7, 5) nanotubes in Fig. 2. Since the emission energies of these peaks were almost identical with the emission energy of the main peak at $E_{22}$ excitation energy of (7, 5) nanotubes, these peaks are also attributed to photon emission from (7, 5) nanotubes. In this paper, we focus on these unassigned PL peaks of (7, 5) nanotubes. Hereafter, we refer to these unassigned peaks as peaks A, B and C, as shown in Fig. 2 and Fig. 3.

Magnifications of the same PLE spectra shown in Fig.2c are presented in Fig.3. In the case of peaks A and C, there was a clear change in the energy differences from the main peak depending on whether the sample was SW$^{13}$CNTs or normal SWNTs, while positions of the main peaks were almost identical. If a certain PL peak is a phonon sideband corresponding to the main PL peak, the amount of the change of the energy difference between the main peak and its sideband should be consistent with the value estimated from the difference of phonon energies confirmed by Raman spectroscopy shown in Fig.1. Dotted lines in Fig.3 (a, b) are the shifted PLE spectra of normal SWNTs by multiplying the mass-ratio factor $\sqrt{12/13}$ to energy difference between excitation photon energy and $E_{ii}$ (i = 1,2 for peak A and C, respectively) transition energy ($E_{exp} - E_{ii}$). Since the positions of peaks A and C were in good agreement with the shifted spectra, these peaks were consistently identified as phonon sidebands corresponding to the main PL peak at $E_{ii}$ energies of (7, 5) nanotubes. Since the excitation energies corresponding to peaks A and C are larger than those of the main PL peaks at $E_{ii}$ energies, phonon emission may be the origin of these peaks. However, the energy shifts from the main PL peaks, 0.21-0.22 eV, are considerably larger than the G-band energy, 0.197 eV. Here, it should be noted that the line shape and energy shift from $E_{ii}$ are remarkably similar to the excitonic phonon sideband predicted by Perebeinos *et al*. [18]. On the other hand, we observed almost no shift of peak B as shown in Fig.3a, indicating that peak B is due to a 'pure electronic' transition without excition-phonon interaction. Further study of the origin of PL peaks such as peak B is in progress by our group.

In conclusion, we have measured PL and Raman spectra of SW$^{13}$CNTs and distinguished excitonic phonon sideband PL peaks from those without exciton-phonon interaction by comparing PLE spectra of SW$^{13}$CNTs and normal SWNTs. With regard to (7, 5) nanotubes, two unassigned PL peaks with energies about 0.21-0.22 eV larger than $E_{11}$ and $E_{22}$ transition energies, respectively, are identified as excitonic phonon sidebands, while the PL peak with an energy about 0.1 eV smaller than the $E_{22}$ transition energy is identified as a 'pure electronic' PL peak without exciton-phonon interaction.

The authors are grateful to E. Einarsson and S. Chiashi (University of Tokyo) for valuable discussions. Parts of this work were financially supported by KAKENHI #16360098 and #13GS0019 from JSPS and MEXT.

**Figure captions**

FIG. 1. Raman spectra measured with a 488nm (2.54 eV) excitation laser and PL maps of normal SWCNTs and SW$^{13}$CNTs. (a) RBM, (b) G-band. Dotted lines are spectra of normal SWNTs shifted by multiplying the Raman shift frequency by the mass ratio factor $\sqrt{12/13}$. PL maps of (c) SWCNTs and (d) SW$^{13}$CNTs.

FIG. 2. PL maps of (a) normal SWNTs and (b) SW$^{13}$CNTs dispersed in surfactant suspension. (c) Comparison of PLE spectra of SW$^{13}$CNTs and normal SWNTs at the emission energy of 1.208 eV (corresponding to the $E_{11}$ energy of (7, 5) nanotubes).

FIG. 3. Comparison of (a) peak A, and (b) peaks B and C in the PLE spectra of normal SWNTs and SW$^{13}$CNTs. Dotted lines for peaks A and C are spectra of normal SWNTs shifted by multiplying the energy difference between the excitation energy and the $E_{ii}$ energy (i =1 for peak A and i =2 for peak C, respectively) by the mass ratio factor $\sqrt{12/13}$.

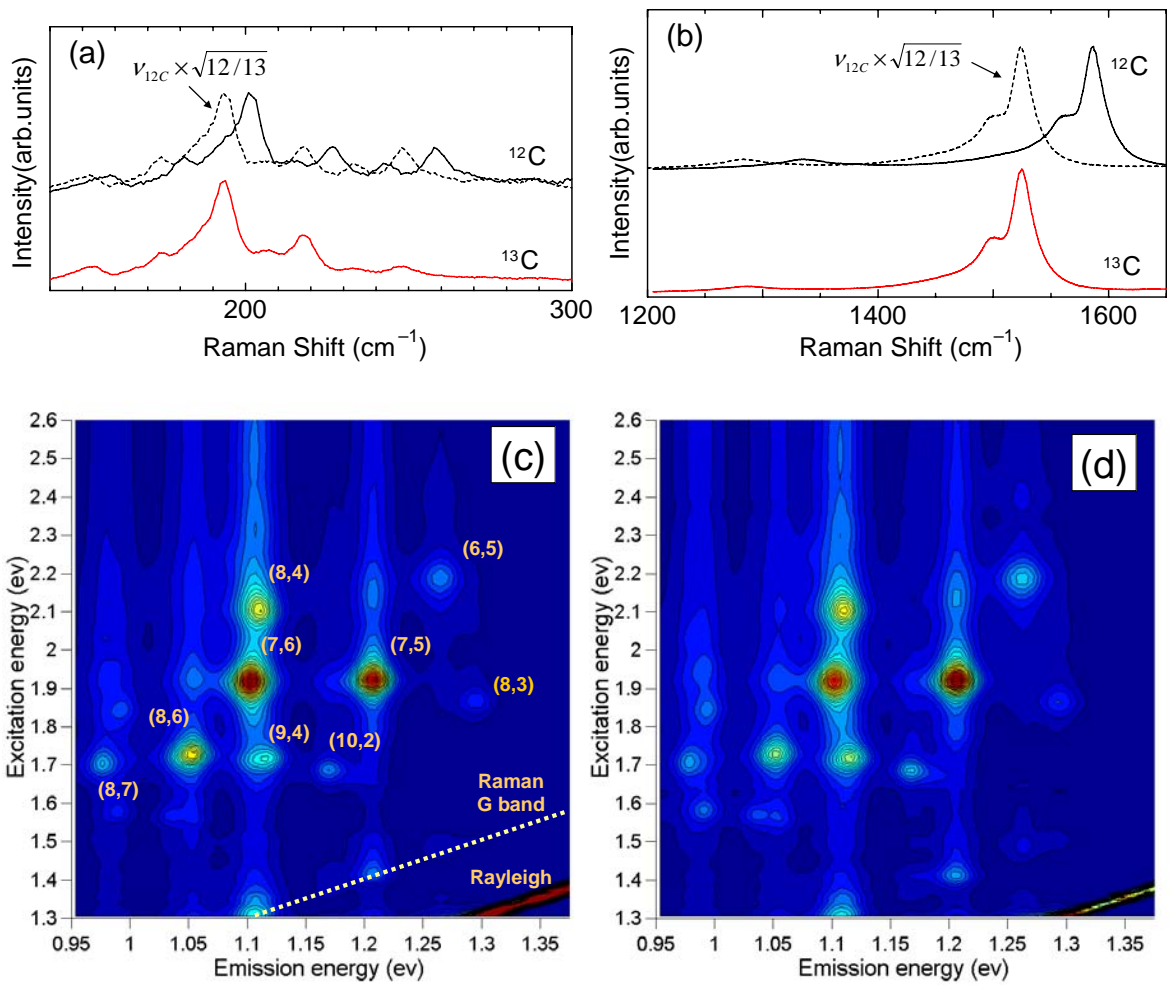

FIG. 1 (color online) Y. Miyauchi *et al.*

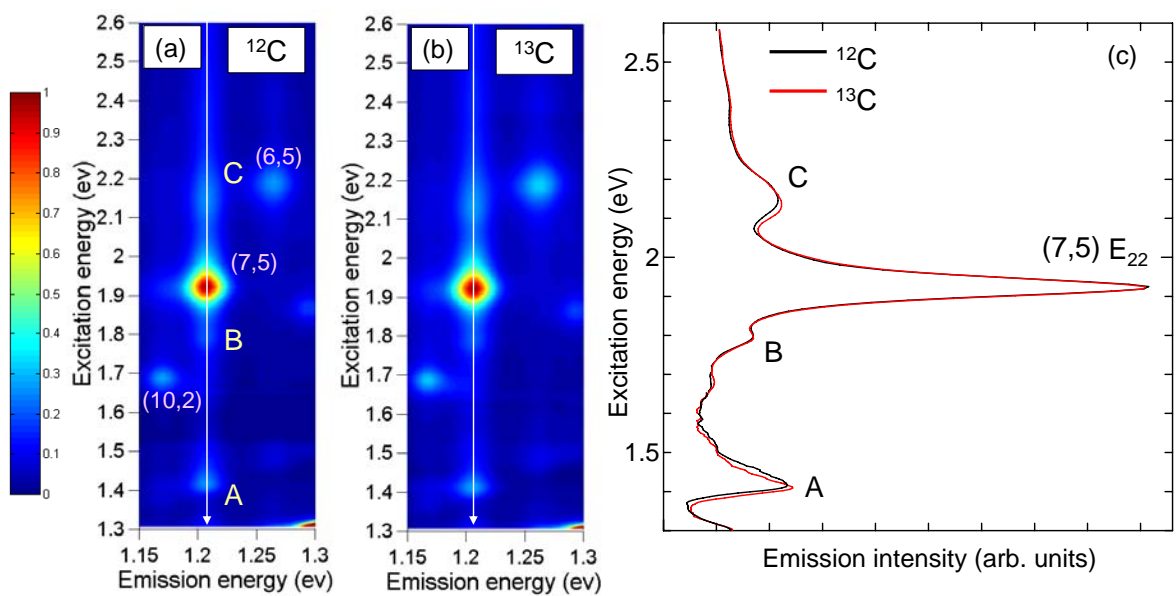

FIG. 2 (color online) Y. Miyauchi *et al.*

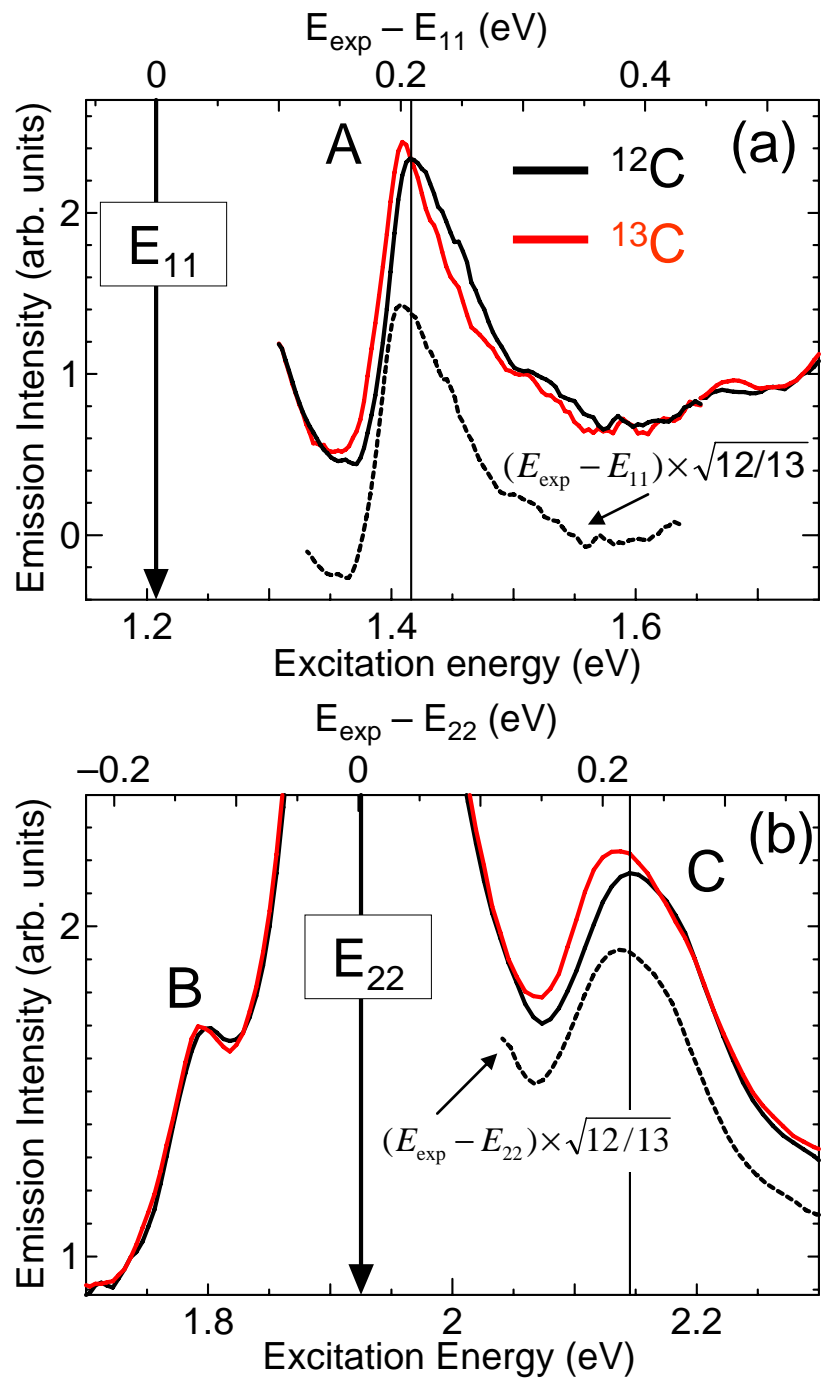

FIG. 3 (color online) Y. Miyauchi *et al.*